\documentstyle[preprint,aps]{revtex}
\input epsf

\begin{document}
\newcommand{\none}{\nonumber\\}
\newcommand{\bea}{\begin{eqnarray}}
\newcommand{\eea}{\end{eqnarray}}
\newcommand{\xx}{\noindent}
\newcommand{\vs}{\vspace*{.5cm}}
\draft


\tighten

\title{ Spontaneous Symmetry Breaking for Scalar QED with Non-minimal
 Chern-Simons Coupling}

\author{ D.S. Irvine${}^{a}$, M.E. Carrington${}^{b,c}$, G.
Kunstatter${}^{c,d}$ and D. Pickering${}^{a}$}

\address{ ${}^a$ Department of Mathematics and Computer Science, Brandon
University, Brandon, Manitoba, R7A 6A9 Canada\\
${}^b$ Department of Physics, Brandon University, Brandon, Manitoba, R7A 6A9
Canada\\
${}^c$  Winnipeg Institute for
Theoretical Physics, Winnipeg, Manitoba, R3B 2E9\\
 ${}^d$ Department of Physics,
 University of Winnipeg, Winnipeg, Manitoba, R3B 2E9 Canada \\
}

\maketitle
\begin{abstract}
We investigate the two-loop effective potential for both minimally and
  non-minimally coupled Maxwell-Chern-Simons theories. The non-minimal
gauge interaction represents the  magnetic moment interaction
between a charged scalar and the electromagnetic field.
In a previous paper we have shown that the two loop effective potential
 for this model is renormalizable with an   appropriate
choice of the non-minimal coupling constant.  We carry out a
 detailed analysis of the
spontaneous symmetry breaking induced by radiative corrections.
As long as the renormalization point for all couplings is chosen to be
the true minimum of the effective potential, both models predict the
presence of spontaneous symmetry breaking.  Two loop corrections are small
compared to the one loop result, and thus the symmetry breaking is
perturbatively stable.
\end{abstract}

\vspace{3ex}

\section{Introduction}

Maxwell-Chern-Simons electrodynamics has been studied extensively in
recent years for a variety of reasons. The Chern-Simons term gives
 the photon a topological mass without spontaneously breaking gauge
 symmetry \cite{refn1}
and allows for the existence of charged particles with
fractional statistics$\cite{refn2}$.
Pure Chern-Simons scalar electrodynamics 
admits topological and non-topological self-dual solitons,
for which many exact solutions to the classical equations
of motion are available$\cite{refn3}$.
 Such theories may have physical significance.
Relativistic three dimensional Chern-Simons theories
provide a consistent  description of
the high temperature limit of four dimensional gauge theories$\cite{refn4}$,
and certain solid state systems with planar dynamics$\cite{refn2}$.
In addition, the non-relativistic version of Maxwell-Chern-Simons theory has
been applied to the fractional Hall effect, and more
recently to rotating superfluid ${}^3$He-A\cite{goryo}.

Work has also been done with a version of scalar electrodynamics in three
dimensions in which
a non-minimal Chern-Simons type gauge interaction is introduced
\cite{ref1,1995}.
The non-minimal coupling in this model represents a
magnetic moment interaction between the charged scalar
and the electromagnetic field. It is of interest for several reasons. Firstly,
it is well known that one of the most important features of scalar
quantum electrodynamics (QED) is
the occurrence of the Coleman-Weinberg mechanism$\cite{refn5}$. In scalar QED
with non-minimal coupling, the Chern-Simons term is generated
through the Coleman-Weinberg mechanism$\cite{ref1}$.  In this sense,
the non-minimal model is the one in which the Chern-Simons term arises
naturally 
rather than being put in by hand.

Another reason that the non-minimal model is of interest involves
the study of vortex solutions.  In recent years, the classical
vortex solutions of 2+1-dimensional
Chern-Simons field theories have received considerable
 attention$\cite{refn3,ref3}$. To find such a solution exactly,
the model must be self-dual. A self-dual theory is one in which
the classical equations of motion can be reduced from second- to
first-order differential equations. 
  In the absence of a Maxwell term, scalar
QED with a Chern-Simons term is self-dual, and the
topological and non-topological
vortex solutions have been found with an appropriately
chosen scalar potential$\cite{refn3}$. However,
if the Maxwell term is present, a self-dual
 Maxwell-Chern-Simons theory can be achieved only if a magnetic
moment interaction between the scalar and the gauge field,
i.e. the non-minimal Chern-Simons coupling,
is introduced$\cite{ref3,ref5}$. 

It is well known that
Maxwell-Chern-Simons scalar QED is renormalizable.
Non-minimal gauge interactions are, however,
notoriously non-renormalizable in four dimensions. There is
some hope that the situation might
be different in three dimensions. Some time ago it was found by
two of us that the non-minimal
Chern-Simons coupling in 2+1-dimensional scalar electrodynamics
is actually renormalizable at the one-loop level$\cite{ref1}$.
 The renormalizability occurs because the non-minimal gauge interaction
contains 
the three-dimensional antisymmetric tensor.  At the two loop level, the full
effective action of the non-minimal theory is not renormalizable.  However, it
has been shown by two of us that the two loop effective potential is
renormalizable providing certain conditions are
satisfied by the coupling constants
of  the model \cite{jeff}.
When the lowest order in the momentum expansion is sufficient, the effective
potential can be used to obtain physical information about spontaneous symmetry
breaking at the two-loop level.

In this paper, we compare the symmetry breaking phase transitions in the
minimal
and non-minimal models at two loop order.
The two-loop behaviour
of the minimal model has recently been analyzed in detail \cite{ref6}.  We
discuss the calculation of the effective potential for this model in  section
II.
Section III contains a brief review of the calculation of the effective
potential for the non-minimal model. Details are given in \cite{jeff}. In
section IV we discuss the renormalization of the minimal effective potential.
In contrast to \cite{ref6} we show that one obtains perturbatively reliable
spontaneous symmetry breaking as long as the minimum of the potential is used
to provide a physical renormalization point.  In section V we discuss the
renormalization of the non-minimal model.  In sections IV and V calculations
are carried out using $Mathematica$. Conclusions are presented in section VI.

\section{Scalar QED with Minimal Chern-Simons Coupling}
\subsection{The Effective Potential}

The minimal model has been studied in a previous paper \cite{ref6}. The
Lagrangian for scalar QED in $2+1$ dimensions with a minimal Chern-Simons
coupling is,
\begin{eqnarray}
{\cal L}=&&\frac{1}{4}F_{\mu\nu}F^{\mu\nu} -
\frac{\kappa}{2}\epsilon^{\mu\nu\rho}A_\mu\partial_\nu A_\rho + {\cal L}_{g.f.}
+ {\cal L}_{F.P.} + \frac{1}{2}|D_\mu \phi|^2 \nonumber \\
&&~~ - \frac{m^2}{2}(\phi^*\phi) - \frac{\lambda}{4!}(\phi^* \phi)^2 -
\frac{\nu}{6!}(\phi^*\phi)^3
\end{eqnarray}
where $D_\mu = \partial_\mu + ieA_\mu$ and $\phi = \chi + i\eta$.
In the $R_\xi$ gauge we have,
\begin{eqnarray}
&&{\cal L}_{g.f.} = -\frac{1}{2\alpha}(\partial_\mu A^\mu - \alpha ev \eta)^2
\\
&& {\cal L}_{F.P.} = - c^\dagger (\partial^2 + \alpha e^2 v \chi) c\,.
\end{eqnarray}

We consider first the effective potential for a theory with a real scalar
field. The effective potential is the energy density of the vacuum in which
the expectation value of the scalar field is given by
$\langle \phi\rangle =v$$\cite{ref7}$.
It can be determined from the effective action
$\Gamma[\tilde{\phi}]$ according to
\begin{eqnarray}
\Gamma[\tilde{\phi}=v]=-(2\pi)^n\delta^{(n)}(0) V_{\rm eff}(v),
\end{eqnarray}
where $\tilde{\phi}$ is the vacuum expectation value of
the scalar field in the presence of the external source.
Using the fact that $\Gamma[\tilde{\phi}]$ is the
generating functional of the proper vertex,
\begin{eqnarray}
\Gamma[\tilde{\phi}]=\sum_{j=1}^{\infty}\frac{1}{j!}
\int d^n x_1d^n x_2{\cdots}d^nx_j \Gamma^{(j)}(x_1,{\cdots},x_j),\nonumber
\end{eqnarray}
one has
\begin{eqnarray}
V_{\rm eff}(v)=-\sum_{j=2}^{\infty}\frac{1}{j!}
\Gamma^{(j)}(0,0,{\cdots},0)v^j,
\label{eq13}
\end{eqnarray}
which  means that one can get the effective potential by calculating
the 1PI vacuum diagrams.  Symmetry breaking occurs when the minimum of the
potential occurs at a non-zero value of $v$.
In our case we have a complex scalar field $\phi = \chi + i\eta$ and we
calculate the effective potential by shifting the real part of the scalar
field, $\chi\rightarrow\chi +v$$\cite{ref7}$.

A regularization scheme must be chosen to handle the
ultraviolet divergences of the theory.
In this paper we shall use dimensional regularization. The use of
dimensional regularization in a theory that explicitly depends on
epsilon tensors involves adopting a complicated form for the gauge
field propagator, as will be discussed below.  In spite of this
complication, dimensional regularization is simpler
than Pauli-Villars regularization because of the fact that it allows us to
preserve explicit gauge symmetry.

There are several problems involved with analytic continuation
to $n$ dimensions.  The first of these is standard. In $n$ dimensional
space-time the mass dimensions
of the fields and parameters are different from their (2+1) dimensional values.
In order to arrange for the  parameters to keep their
original mass dimensions, one introduces a mass scale $\mu$ in such a way that,
for each parameter, the dimensional changes are absorbed into a factor
$\mu^{(3-n)a}$ where $a$ is a number that depends on the original mass
dimensions of the parameter.  The second problem is more complicated.
Dimensional regularization
in a theory with a three-dimensional antisymmetric tensor
$\epsilon_{\mu\nu\rho}$ must be handled carefully. It has been
explicitly shown that naive dimensional regularization schemes
cannot make the theory well defined when they are applied
to a Chern-Simons type model$\cite{ref9}$. Therefore,
in carrying out dimensional regularization we must adopt
the three-dimensional analogue of the consistent definition for
$\gamma_5$, which was originally proposed by 't Hooft and
Veltman$\cite{ref10}$, and later given a strict mathematical justification by
Breitenlohner and Maison$\cite{ref11}$. The explicit definition of
this dimensional continuation for a Chern-Simons-type theory was
explained in Ref.$\cite{ref12}$ where it is shown that this regularization
method is indeed compatible with the Slavnov-Taylor identities.

We express the result for the effective potential in terms of the following
masses.  The scalar masses are,
\bea
&&m_\chi^2(v) = m^2 + \frac{\lambda}{2} v^2 + \frac{\nu}{24} v^4 \nonumber\\
&& m_\eta^2(v) = m^2 + \frac{\lambda}{6} v^2 + \frac{v}{120} v^4
\eea
and the gauge boson masses are
\bea
&&m_{1,2}(v) = \frac{1}{2} \left\{ \sqrt{\kappa^2 + 4(ev)^2} \pm
|\kappa|\right\}\\
&& m^2_3(v):= m_1(v) m_2(v) \nonumber
\eea
The effective potential can be written in three pieces \cite{ref6}:
\bea
V_{eff}(v) = V_{tree} + V_{1-loop} + V_{2-loop}
\eea
with
\bea
V_{tree} = \frac{\nu}{6\!}v^6
\eea
\bea
V_{1-loop} = -\frac{\hbar}{12\pi}\left\{ m_\chi^3(v) + m_\eta^3(v) + m_1^3(v) +
m_2^3(v) \right\}
\eea
\bea
V_{2-loop} = V_{q1} + V_{q2} + V_{c1} + V_{c2} + V_{c3} + V_{c4}
\eea
\bea
V_{q1} = \frac{\hbar^2}{(4\pi)^2}\left\{ 3\left(\frac{\lambda}{4\!} + \frac{15
\nu v^2}{6\!} \right) m_\chi^2 + 3\left( \frac{\lambda}{4!} + \frac{3\nu
v^2}{6\!}\right) m_\eta^2 + 2\left( \frac{\lambda}{4\!} + \frac{9 \nu
v^2}{6\!}\right) m_\chi m_\eta \right\}
\eea
\bea
V_{q2} = \frac{e^2 \hbar^2 \mu^{2(n-3)}}{16\pi^2}\frac{(m_\chi + m_\eta) (m_1^2
+ m_2^2)}{m_1 + m_2}
\eea
\bea
V_{c1} =&& -\hbar^2\left[3\left(\frac{\lambda}{3\!} v + \frac{\nu}{36}
v^3\right)^3 + \left( \frac{\lambda}{3\!} v + \frac{\nu}{60} v^3\right)^2
\right] I_{div}\nonumber\\
&&+\frac{\hbar^2}{16\pi^2}\left\{ 3 \left( \frac{\lambda}{3\!} v +
\frac{\nu}{36}v^3\right)^2 {\rm ln} \frac{3m_\chi}{\mu} +
\left(\frac{\lambda}{3\!} v + \frac{\nu}{60} v^3\right)^2 {\rm ln}\frac{m_\chi
+ 2m_\eta}{\mu}\right\}
\eea
\bea
V_{c2} &&= \frac{e^2 \hbar^2}{2a} \left[ 2(m_\chi^2 + m_\eta^2) - (m_1+m_2)^2 +
3m_3^2\right]I_{div} \\
&&+\frac{e^2 \hbar^2 }{32\pi^2 a} \left[\right. \left[ m_\chi m_\eta -
\frac{(m_\chi + m_\eta)(2[m_\chi-m_\eta)^2 + m_1^2 + m_2^2]}{m_1+m_2}\right]
-\frac{(m_\chi^2 - m_\eta^2)^2}{m_3^2} \ln \frac{m_\chi + m_\eta}{\mu}
\nonumber \\
&& -\sum_{x = 1,2} \frac{2m_x^2(m_\chi^2 + m_\eta^2) - m_x^4 -(m_\chi^2 -
m_\eta^2)^2}{m_x(m_1 + m_2)} \ln \frac{m_x + m_\chi + m_\eta}{\mu} -
\frac{5}{12} \frac{\kappa^2}{a^2}\left.\right] \nonumber
\eea
\bea
V_{c3} =&& -\frac{3 \hbar^2 e^4 v^2}{2a^2} I_{div} - \frac{\hbar^2 e^4 v^2
}{32\pi^2 a^2}\left[3-\frac{2m_\chi}{m_1+m_2} - \frac{2(m_\chi^2 +
6m_3^2)}{(m_1+m_2)^2}\right] \\
&&+\frac{\hbar^2 e^4 v^2}{64 \pi^2 a^2}\left[\right.\frac{2[(m_1-m_2)^2 -
m_\chi^2]^2}{m_3^2(m_1+m_2)^2} \ln \frac{m_1+m_2+m_\chi}{\mu} +
\frac{m_\chi^4}{m_3^4}\ln \frac{m_1}{\mu} \nonumber \\
&&+\sum_{x=1,2}\left[\frac{(4m_x^2 - m_\chi^2)^2}{m_x^2(m_1+m_2)^2}\ln
\frac{2m_x + m_\chi}{\mu} - \frac{2(m_x^2 - m_\chi^2)^2}{m_3^2
m_x(m_1+m_2)}\ln\frac{m_x + m_\chi}{\mu}\right]\left.\right] \nonumber
\eea
where
\bea
I_{div} = \frac{1}{32\pi^2}\left\{ \frac{1}{3-n} -\gamma + 1 + {\rm ln}
4\pi\right\}
\eea

\section{Scalar QED with Non-minimal Chern-Simons Coupling}

The Lagrangian for scalar QED in $2+1$ dimensions with a non-minimal
Chern-Simons coupling is$\cite{ref1}$
\begin{eqnarray}
{\cal L}=\frac{1}{2}(D_{\mu}\phi)^*D^{\mu}\phi-\frac{1}{4}F_{\mu\nu}F^{\mu\nu}
-\frac{i}{8}\gamma_0\epsilon^{\mu\nu\rho}F_{\nu\rho}\left[\phi^*D_{\mu}\phi-
(D_{\mu}\phi)^*\phi\right] - \frac{\lambda}{6!}(\phi^*\phi)^3
\label{eq1}
\end{eqnarray}
The non-minimal coupling is similar to the minimal Chern-Simons term in several
ways: it is odd under parity reversal, topological (in the sense that it does
not depend on the space-time metric) and gauge invariant. In spite of the fact
that the theory is super non-renormalizable as a consequence of the presence of
a coupling constant ($\gamma_0$) which has negative mass dimension, it can be
shown that the Lagrangian is renormalizable to one loop order \cite{ref1} and
that the effective potential is renormalizable to two loop order when certain
constraints on the couplings are satisfied \cite{jeff}.  Note that a
$(\phi^*\phi)^2$ term has not appeared in the Lagrangian since a term of this
form would make a super renormalizable contribution.  This term can appear in
the renormalized theory however, since the Lagrangian doesn't possess a
symmetry that guarantees its vanishing.

We define the following masses:
\bea
&& m_{\chi}^2=\frac{\lambda}{4!}v^4 \none
&& m_{\eta}^2=\frac{\lambda}{5!}v^4 \none
&& m_1=\frac{1}{2}e|v|\left(\sqrt{(\gamma_0v)^2+4}+\gamma_0v\right),
\nonumber\\
&& m_2 = \frac{1}{2}e|v|\left(\sqrt{(\gamma_0v)^2+4}-\gamma_0v\right),
\label{masses}\\
&& m_3^2 = m_1 m_2=  e^2 v^2 \nonumber
\eea
and use the identities
\begin{eqnarray}
&&e^2 = \frac{1}{v^2}m_3^2, \nonumber \\
&& e\gamma_0 = \frac{1}{v^2} (m_1-m_2), \\
&&\gamma_0^2 = \frac{1}{v^2}\frac{(m_1-m_2)^2}{m_1m_2}\,. \nonumber
\end{eqnarray}

The result for the effective potential is
\bea
V(v) = V_{tree} + V_{1-loop} + V_{2-loop}
\eea
\bea
V_{tree} = \frac{\lambda}{6!} v^6
\eea
\begin{eqnarray}
V_{1-loop}&=&-\frac{\hbar}{12\pi}\left[m_{\chi}^3
+m_{\eta}^3+m_1^2 + m_2^2\right]
\end{eqnarray}
\bea
V_{2-loop} = V_1 + V_2 + V_3 + V_4 + V_5
\eea
\bea
V_1 = \frac{\hbar^2\lambda v^2}{16\pi^2}\left(\frac{3}{2}m_{\chi}^2+
\frac{3}{10}m_{\eta}^2+\frac{1}{10}m_{\chi}m_{\eta} \right)
\eea
\bea
V_2 = \frac{\hbar^2}{4\pi^2}\frac{1}{v^2}\frac{m_\chi+m_\eta}{m_1+m_2}
\left[(m_1^2-m_2^2)^2+2 m_1^2 m_2^2\right]
\eea
\bea
V_3 = &=& \frac{\hbar^2}{2}(\lambda v^3)^2
\left(\frac{1}{3!^3}+\frac{1}{(5{\times}3!)^2{\times}2!}
\right)\frac{1}{32\pi^2}
\left(\frac{1}{3-n}-\gamma+1+\ln 4\pi\right)\nonumber\\
&&-\frac{\hbar^2}{2}(\lambda v^3)^2 \frac{1}{32\pi^2}
\left[\frac{1}{3!^3}\ln\frac{9m_\chi^2}{\mu^2}
+\frac{1}{(5{\times}3!)^2{\times}2!}\ln\frac{(2m_{\eta}+m_\chi)^2}{\mu^2}
\right]
\eea
\bea
V_4 = && V_4^{div} + V_4^{finite} \nonumber \\
V_4^{div} = &&
 ={\hbar^2\over 64\pi^2}\left({1\over 3-n} -\gamma
+1 +\ln(4\pi)\right)\left(\left({4\over v^2}+\gamma_0^2\right)
\left({1\over2} \left((m_1-m_2)^2 + m_1m_2\right)(m_\chi^2+m_\eta^2)
\right.\right.
\none
& &-{1\over 4}(m_1^2-m_2^2)^2+\left.\left.{1\over 4}m_1m_2(m_1-m_2)^2 - {1\over
4}m_1^2m_2^2\right)
 - {1\over 4} \gamma_0^2 (m_\chi^2-m_\eta^2)^2\right)
\none
V_4^{finite} =&&
-{\hbar^2\over 64 \pi^2 v^2} (m_\chi-m_\eta)^2
\ln\left({(m_\chi^2+m_\eta^2)^2\over \mu^2}\right)\none
& &+{\hbar^2(\gamma_0^2v^2+4)m_2\over 256
\pi^2v^2(m_1+m_2)}((m_\chi+m_\eta)^2-m_2^2)((m_\chi-m_\eta)^2-m_2^2)
\ln\left({(m_\chi+m_2+m_\eta)^2\over \mu^2}\right)\none
& &+{\hbar^2(\gamma_0^2v^2+4)m_1\over 256
\pi^2v^2(m_1+m_2)}((m_\chi+m_\eta)^2-m_1^2)((m_\chi-m_\eta)^2-m_1^2)
\ln\left({(m_\chi+m_1+m_\eta)^2\over \mu^2}\right)\none
& & -{\hbar^2 (\gamma_0^2v^2+4)\over 1536 \pi^2 v^2(m_1+m_2)}\left(
 12(m_1^2+m_2^2)(m_\chi^3-m_\chi^2m_\eta-m_\eta^2m_\chi+m_\eta^3)
+5m_1^5+5m_2^5
\right.\none
& &-(m_1^3+m_2^3)(10m_\chi^2+12m_\chi
 m_\eta + 10 m_\eta^2)+m_3^2(m_1+m_2)(10m_\chi^2+10m_\eta^2-5m_3^2)
\none
& & \left. +12(m_1^4+m_2^4)(m_\chi+m_\eta)\right)\none
& &-{\hbar^2\gamma_0^2\over 7680 \pi^2}
\left( 6m_1^4+6m_2^4 - (12
m_\eta^2+12m_\chi^2+6m_3^2)(m_1^2+m_2^2)  + 25 m_3^4\right.
\none
& &\left. -26(m_\chi^2+m_\eta^2)m_3^2 + 60m_\eta
  m_\chi(m_\chi^2+m_\eta^2) + 25(m_\chi^2-m_\eta^2)^2\right).
\end{eqnarray}
\bea
V_5 = &&V_5^{e^2} + V_5^{e\gamma_0} + V_5^{\gamma_0^2} \none
V_5^{e^2} = &&
-{3\hbar^2\over 16\pi^2 v^2}\left({1\over 3-n} -\gamma
+1 +\ln(4\pi)\right) \mu^{(2n-6)} m_3^4 +{\hbar^2 m_\chi^4\over32 \pi^2 v_2}
\ln\left({m\chi^2\over
     \mu^2}\right) \none
& &
  -{\hbar^2m_1(m_2^2-m_\chi^2)^2\over 16\pi^2 (m_1+m_2) v^2}
   \ln\left({(m_2+m_\chi)^2\over \mu^2}\right) -{\hbar^2
m_2(m_1^2-m_\chi^2)^2\over 16 \pi^2(m_1+m_2)v^2}
    \ln\left({(m_1+m_\chi)^2\over \mu^2}\right)
\none
 & & +{\hbar^2m_1^2(4m_2^2-m_\chi^2)^2\over 32 (m_1+m_2)^2 \pi^2
   v^2}\ln\left( {(2m_2+m_\chi)^2\over \mu^2}\right)
+{\hbar^2m_2^2(4m_1^2-m_\chi^2)^2\over 32 (m_1+m_2)^2 \pi^2
   v^2}\ln\left( {(2m_1+m_\chi)^2\over \mu^2}\right)\none
& &+{\hbar^2 m_3^2 ((m_1-m_2)^2-m_\chi^2)^2\over 16(m_1+m_2)^2\pi^2 v^2}
   \ln\left({(m_1+m_2+m_\chi)^2\over \mu^2}\right)\none
& & +{\hbar^2 m_3^4(-3(m_1-m_2)^2 + 2m_\chi(m_1+m_2+m_\chi))\over
   8 (m_1+m_2)^2 \pi^2 v^2}\label{I} \none
V_5^{e\gamma_0} = && -{5\over 8\pi^2 v^2} \hbar^2 \left({1\over 3-n} -\gamma
+1 +\ln(4\pi)\right) \mu^{(2n-6)}(m_1-m_2)^2m_3^2\none
& &-{\hbar^2 m_3^2 (m_1-m_2)^2 (49m_1^2 + 49m_2^2 - 42(m_1+m_2) m_\chi +
  26 m_3^2 - 12 m_\chi^2)
\over 48 (m_1+m_2)^2 \pi^2 v^2}\none
& & -{ \hbar^2(4 m_1^2-m_\chi^2)^2 (m_2-m_1)m_2\over 16(m_1+m_2)^2\pi^2
  v^2}
   \ln\left({(2m_1+m_\chi)^2\over \mu^2}\right)\none
& & + { \hbar^2(4 m_2^2-m_\chi^2)^2 (m_2-m_1)m_1\over 16(m_1+m_2)^2\pi^2
  v^2}
   \ln\left({(2m_2+m_\chi)^2\over \mu^2}\right)\none
& & +{\hbar^2(m_1-m_2)^2((m_1-m_2)^2-m_\chi^2)^2\over
    16(m_1+m_2)^2\pi^2 v^2}
    \ln\left({(m_1+m_2+m_\chi)^2\over \mu^2}\right)\none
& & +{\hbar^2(m_1-m_2)(m_2^2-m_\chi^2)^2 \over 16\pi^2(m_1+m_2)v^2}
  \ln\left({(m_2+m_\chi)^2\over \mu^2}\right)\none
& & -{\hbar^2(m_1-m_2)(m_1^2-m_\chi^2)^2 \over 16\pi^2(m_1+m_2)v^2}
  \ln\left({(m_1+m_\chi)^2\over \mu^2}\right)\label{II}\none
V_5^{\gamma_0^2} = &&  -{\hbar^2(m_1-m_2)^2\over 64 \pi^2 v^2}\left[
\mu^{(2n-6)}  \left({1\over 3-n} -\gamma
+1 +\ln(4\pi)\right)(25(m_1^2+m_2^2)-35m_3^2-6m_\chi^2)\right.
\none
& &-{2(4m_1^2-m_\chi^2)^2\over (m_1+m_2)^2}
    \ln\left({(2m_1+m_\chi)^2\over \mu^2}\right)-{2(4m_2^2-m_\chi^2)^2\over
(m_1+m_2)^2}
    \ln\left({(2m_2+m_\chi)^2\over \mu^2}\right)\none
& &+{(m_1^2-m_\chi^2)^2 m_1\over (m_1+m_2)m_3^2}
\ln\left({(m_1+m_\chi)^2\over \mu^2}\right)+{(m_2^2-m_\chi^2)^2 m_2\over
(m_1+m_2)m_3^2}
\ln\left({(m_2+m_\chi)^2\over \mu^2}\right)\none
& &-\left.{(m_1-m_2)^2((m_1-m_2)^2-m_\chi^2)^2\over (m_1+m_2)^2m_1m_2}
\ln\left({(m_1+m_2+m_\chi)^2\over \mu^2}\right)\right.\none
& & +{1\over 2310 (m_1+m_2)^2}
  \left( 89981(m_1^4+m_2^4) - 110880m_\chi(m_1^3+m_2^3)
\right.\none
& & -(3837m_3^2+28158m_\chi^2)(m_1^2+m_2^2)
+(-46200m_\chi m_3^2+9240m_\chi^3)(m_1+m_2)\none
& & \left.\left.-19356m_\chi^2m_3^2
 +34124m_3^4\right)\right]
\label{III}
\eea
It is straightforward to show that $V_4^{div}$ contains terms that are
proportional to $v^8$ \cite{jeff}.  As will be discussed in the next section,
these terms must be removed for the potential to be renormalizable. We achieve
this by imposing the condition
\bea
\gamma_0^2 = \frac{(3\pm \sqrt{5})\lambda}{60 e^2}
\label{gl}
\eea
In our analysis, we choose $\gamma_0 = \sqrt{(3+\sqrt{5})\lambda/60e^2}$.

\section{Analysis and Results for the Minimal Model}

The divergent terms in the effective potential are handled through the
renormalization process.  They are effectively `absorbed' by redefining the
physical parameters.  We throw away the divergent terms, and replace them by a
set of counterterms of the form,
\bea
V_{ct} = A\frac{v^2}{2} + B\frac{v^4}{4\!} + C\frac{v^6}{6\!}
\eea
The constants $A$, $B$, and $C$ are determined, as functions of the parameters
of the theory, by imposing the renormalization conditions.  A typical choice
for these conditions is:
\bea
&&V|_{v=0} = 0 \nonumber\\
&& \frac{\partial^2V}{\partial v^2}|_{v=0} = m^2  \nonumber\\
&& \frac{\partial^4V}{\partial v^4}|_{v=0} = \lambda  \label{rcond1}\\
&& \frac{\partial^6V}{\partial v^6}|_{v=\sqrt{M}} =  \nu  \nonumber
\eea
This choice of renormalization conditions is made in \cite{ref6} and is
conventional because the algebra is simpler when $v$ is set to zero in the
expressions for the derivatives (the sixth derivative cannot be evaluated at
$v=0$ because it is singular there).

However, in order to ensure that perturbation theory will be reliable, it is
advantageous to choose renormalization conditions that give rise to the
`physical' parameters. Towards this end, we choose renormalization conditions
of the form,
\bea
&&V|_{v=0} = 0 \nonumber\\
&& \frac{\partial^2V}{\partial v^2}|_{v=\sqrt{M}} = m^2  \nonumber\\
&& \frac{\partial^4V}{\partial v^4}|_{v=\sqrt{M}} = \lambda  \label{rcond2}\\
&& \frac{\partial^6V}{\partial v^6}|_{v=\sqrt{M}} =  \nu  \nonumber
\eea
The physical parameters are the ones for which the renormalization point (the
point at which the derivatives are evaluated) is  the position of the global
minimum.  The constraint
\bea
\frac{\partial V}{\partial v}|_{v=\sqrt{M}} = 0
\label{derconst}
\eea
 ensures that the potential has an extremum at $v=\sqrt{M}$ but not necessarily
a global minimum (this extrema could be a local minimum, or a maximim, or an
inflection point). To have spontaneous symmetry breaking it must be possible to
choose a set of parameters $M$, $m$, $\lambda$ and $\nu$ so that the couplings
are small and the resulting effective  potential has a global minimum at
$v=\sqrt{M}$.
 The constraint (\ref{derconst}) will give us a condition on the couplings of
the form $F(M, m,\lambda,\kappa, \nu) =0$ which we can use to implicitly
determine  one of the parameters: we will determine $\kappa$ as a function of
the other variables.
The parameters defined in this way are physical in the sense that they
correspond to the parameters of the fundamental excitation that lives in the
bottom of the potential well and interacts with the rest of the world through
perturbation theory.

If the renormalization point is not chosen to be the position of the global
minimum than the parameters defined by (\ref{rcond2}) and (\ref{derconst}) will
not be the physical ones.  If these `unphysical' parameters are close to the
physical ones then perturbation theory will give approximately the same answer
with either choice for the parameters.  (If we calculate to all orders in PT
then the result will be the same, even if the couplings used are far away from
the physical parameters.  This is equivalent to saying that the physics is
independent of the choice of $M$.  However, different choices of $M$ are not
equally useful, since in pratice we cannot calculate to all orders in
perturbation theory).

  Notice that the logic of the renormalization process seems to be exactly
backwards: the values of the couplings (or the values of the derivatives at the
position of the minima) contain physical information, since these couplings
define the shape of the potential at the global minimum.  But it is just these
couplings that we are choosing. The catch is that we restrict ourselves to
choices of couplings that correspond to symmetry breaking. More precisely, we
proceed as follows: we start with the physical renormalization conditions
(\ref{rcond2}) and choose some values for $M$, $m$, $\lambda$ and $\nu$.  We
determine $\kappa$ from the derivative constraint (\ref{derconst}).  (Note that
it is equivalent to say that we choose the physical parameters  $\kappa$, $m$,
$\lambda$ and $\nu$ and find the corresponding $M$).  If it is possible to
choose a set of parameters $M$, $m$, $\lambda$ and $\nu$ so that the couplings
are small and the resulting effective  potential has a global minimum at
$v=\sqrt{M}$, then the effective potential gives a perturbatively reliable
description of spontaneous symmetry breaking.

We use units in which $\hbar=1$.   We set $e=1$ or effectively we rescale
parameters to obtain dimensionless variables: $\tilde v = v/e$, $\tilde M =
M/e^2$, $\tilde m = m/e^2$, $\tilde \lambda = \lambda /e^2$ and $\tilde \kappa
= \kappa /e^2$ and suppress the tildes.
We choose values for $M$, $\nu$, $m$ and $\lambda$.  We use (\ref{rcond2}) and
the value of $\kappa$  determined by the derivative constraint
(\ref{derconst}).  Using the following values,
\bea
M=1\,;~~~~\nu=0.0005\,;~~~~m=.2709\,;~~~~\lambda = .2200 \label{MMMv}
\eea
we obtain two values of $\kappa$ from the derivative constraint:
$\kappa=\{1.6428,~20.7494\}$.  The resulting potential is very insensitive to
the value of $\kappa$.  We use the larger value and obtain the two loop
potential shown in Fig. [1] which has a global minimum at $v=\sqrt{M}$.

Under certain circumstances, it is advantageous to  use the renormalization
conditions (\ref{rcond1}) instead of (\ref{rcond2}) because the choice
(\ref{rcond1}) is simpler algebraically.  However, as discussed above,
perturbative calculations using the renormalization conditions (\ref{rcond1})
are in general
less reliable because the  couplings differ from the physical values used
in the renormalization conditions (\ref{rcond2}) and (\ref{derconst}). As an
example, we look at the two loop potential obtained using the renormalization
conditions (\ref{rcond1}) with $m=0$, $\lambda=0$ and $\nu=.0005$.  We impose
the constraint (\ref{derconst}) to obtain the complete set of parameters,
\bea
M=1\,;~~~~\nu=0.0005\,;~~~~m=\lambda=0\,;~~~~\kappa = \{.95757,~~25.1850\}
\label{00Mv}
\eea
Using the larger value of $\kappa$ we obtain a potential with a global minimum
at $v=\sqrt{M}$ as shown in Fig. [2].
We now calculate the values of the second and  fourth derivatives at $v=\sqrt{M}$ for the
potential in Fig. [2] and obtain respectively $1.014 \times 10^{-6}~$ and
$5.464 \times 10^{-5},$ which of course
differ from the values at the origin. To check the consistency of
the calculation,  we calculate $\kappa$  using these values with 
the physical renormalization conditions (\ref{rcond2}).  The result is
25.2992 which demonstrates good agreement with the value obtained from the
renormalization conditions (\ref{rcond1}) and shows that our numerical
calculations are reliable to three digits in this case. Again, we stress that only
the values of the derivatives at the true minimum are physical. Moreover,  the
renormalization conditions (\ref{rcond1}) are imposed at $v=0$ which is in
general far from the true minimum and therefore in a region where the  perturbative expansion
may not be reliable. We will return to this point later.

In \cite{ref6} it is claimed that the symmetry breaking disappears when $M$ is
changed at the one loop level.  The authors then claim that this indicates that
perturbation theory is breaking down, and that it is necessary to go beyond one
loop. We have shown that this breakdown of perturbation theory occurs when
non-physical values of the parameters are used.  Fig. [3] shows the one loop
effective potential renormalized using (\ref{rcond1}) with $m=\lambda=0$,
$\kappa=20$, $\nu=0.0005$ and $M = \{.5,~1,~5\}$. The minima for these graphs
are not at $v=\sqrt{M}$ and there is no symmetry breaking for $M=.5$.
Fig. [4] shows the one loop potential for the same range of $M$ values using
the renormalization conditions (\ref{rcond2}) and (\ref{derconst}). The
parameters for the three curves are:
\bea
&& M=.5\,,~m=.2709\,,~\lambda=.2188\,,~\nu=0.0005\,,~\kappa=1.5588 \nonumber \\
&&
M=1\,,~m=.2709\,,~\lambda=.2188\,,~\nu=0.0005\,,~\kappa=\{1.97484\,~14.2495\}
\nonumber \\
&& M=5\,,~m=.2709\,,~\lambda=.2188\,,~\nu=0.0005\,,~\kappa=3.4534 \nonumber
\eea
In the case where the derivative constraint has two solutions, we choose the
smaller value of $\kappa$. As shown in Fig. [4], symmetry breaking persists
when the physical parameters are used.  Thus we have shown that the result
obtained by the authors of \cite{ref6}, that the symmetry breaking depends on
the choice of $M$, is an artifact of their choice of renormalization
conditions.

 Fig. [5] shows the one and two loop potentials for $M=1$ obtained using
(\ref{rcond2}) and (\ref{derconst}).  On the same graph we have also plotted an
expanded expression for the two loop potential.  This expansion is useful to
clarify the physical content of the potential because the two loop expression
is so messy.  The expansion parameter is $z=v^2/\kappa^2$.  As long as the
minimum occurs at a value of $v$ that corresponds to a small value of $z$, the
expanded potential will reliably describe the symmetry breaking.  We will show
that this expansion is useful under certain conditions: If the renormalization
conditions (\ref{rcond1}) are used, with the choices $m=\lambda=0$, the
expansion corresponds to an approximate Coleman-Weinberg limit in the sense
that the constraint (\ref{derconst}) takes the form of a dimensional
transmutation relation which allows us to rewrite the dimensionless parameter
$\nu$ in terms of the parameter $\kappa$.
Using the renormalization conditions (\ref{rcond1}) with $m=\lambda=0$ the
expanded two loop potential has the form,
\bea
V(v) = \frac{\nu v^6}{6\!} + D v^6\left({\rm ln} \frac{v}{\sqrt{M}} -
\frac{49}{20}\right)
\eea
with
\bea
D= \frac{1}{32 \pi^2}\left( 16\frac{1}{\kappa^4} - \frac{11}{30}
\frac{\nu}{\kappa^2} + \frac{7}{675} \nu^2 \right) \label{D}
\eea
Imposing the derivative constraint gives the condition
\bea
\frac{\nu}{5!}  -\frac{137}{10} D =0 \label{eqnD}
\eea
When
\bea
\nu = {\cal O} \left(\frac{1}{\kappa^4}\right)
\eea
equations (\ref{D}) and (\ref{eqnD}) give,
\bea
\nu = \frac{822}{\pi^2} \frac{1}{\kappa^4}
\eea
This expression is a dimensional transmutation relation.
In Fig. [5] we use the parameters $M=1$, $m=.2709$, $\lambda=.2188$ and
$\nu=0.0005$.  For the one loop potential we obtain
$\{\kappa=1.97484,~14.2495\}$; for the two loop potential we have
$\{\kappa=1.64451,~14.9877\}$; and for the expanded potential $\kappa=10.6287$.
 We use the large value of $\kappa$ in each case, so that the expansion
parameter $z=v^2/\kappa^2$ is small for $v$ in the vicinity of the global
minimum.  We note that  the two loop potential is extremely sensitive to the
value of $m$, and the expanded potential is sensitive to $\lambda$. If we move
away from the values chosen above, either the value of $\kappa$ changes
drastically, or the derivative constraint (\ref{derconst}) has no solution.
The figure shows that the three curves are almost identical in the vicinity of
the minimum.  At large $v$  the curves separate.  There is a global minimum at
$v=\sqrt{M}$ and thus we have established that, for this choice of parameters,
symmetry breaking exists and is perturbatively reliable.

\section{Analysis and Results for the Non-Minimal Model}

When working with the non-minimal model we must impose the constraint
(\ref{gl}).  This condition between the two parameters $\lambda$ and $\gamma_0$
removes the $v^8$ terms in the effective potential and is therefore necessary
to obtain a renormalizable result.
For the non-minimal model we must use renormalization conditions of the same
form as those in (\ref{rcond2}) because the derivatives of the non-minimal
potential are divergent at $v=0$. We use
\bea
&&V|_{v=0} = 0 \nonumber\\
&& \frac{\partial^2V}{\partial v^2}|_{v=\sqrt{M}} = m^2  \nonumber\\
&& \frac{\partial^4V}{\partial v^4}|_{v=\sqrt{M}} = \tau  \label{rcond2b}\\
&& \frac{\partial^6V}{\partial v^6}|_{v=\sqrt{M}} =  \lambda  \nonumber
\eea
Note that we have allowed for the presence of a term of the form $\tau v^4  /
4!$ in the renormalized effective potential even though the corresponding term
does not appear in the bare Lagrangian.  As discussed earlier, a term of this
form can  appear in  the renormalized theory, since the Lagrangian doesn't
possess a symmetry that guarantees its vanishing.

We obtain physical couplings by imposing the constraint (\ref{derconst}) which
gives an equation of the form $f(m,\tau,\lambda,M)=0$.  We find solutions to
this equation by looking at a range of $M$'s.  For each $M$ we choose values
for $m$ and $\tau$ which permit physical solutions to the derivative
constraint: values of $\lambda$ which are real and less than one.   We further
restrict ourselves to  $\lambda$'s which give an effective potential that
approaches infinity when $v\rightarrow \infty$ and exhibits symmetry breaking
with a global minimum at $v=\sqrt{M}$.

The algebraic expressions for the effective potential are much more complicated
for the non-minimal model than for the minimal model. We use an expanded
expression for the two loop effective potential to make it easier to identify
the choices of parameters that lead to symmetry breaking, and to check that
reliable results are produced for the full two loop effective potential.  The
expansion parameter is $\epsilon = v \gamma_0$. This parameter is small in the
vicinity of the minimum for all of the examples we look at, as can be seen from
Table [1].  In contrast to the case of the minimal model, the expansion does
not produce a Coleman-Weinberg type dimensional transmutation relation.  To
understand this point, we remind ourselves how this relation was obtained for
the minimal model.  The first step was to expand the potential in  the
parameter $z=v^2/\kappa^2$.  When we imposed (\ref{derconst}) on the expanded
potential we obtained a result in which all of the logarithm terms disappeared.
Furthermore, the variable $M$ dropped out. We were left with a constraint of
the form $F(m,\lambda,\nu,\kappa)=0$.  To further simplify, we choose
$m=\lambda=0$ and obtained a constraint between $\nu$ and $\kappa$. This
constraint is the dimensional transmutation relation for the minimal model.

Now compare what happens with the non-minimal model.  If we expand in the
variable $\epsilon = v \gamma_0$ we obtain a constraint from (\ref{derconst})
in which the logarithms have disappeared.  However, the parameter $M$ does not
drop out, and there is one less coupling than in the minimal case because of
the constraint (\ref{gl}).  We are left with an equation of the form,
$f(m,\tau,\lambda,M)=0$. If we set $m=\tau=0$ we find that $\lambda$ is
determined for each choice of $M$.

We compare the one loop, the two loop and the expanded version of the two loop
potential, for several values of $M$. The results are shown in Table [1] and
Figs. [6-9]. For $M=.5$ no results are given for the expanded potential because
in this case the constraint (\ref{derconst}) has no solutions for  $\lambda$
positive and less than one.  The figures show that symmetry breaking exists and
perturbation theory is reliable.

\vskip .4in
\begin{center}
\begin{tabular}{l@{\extracolsep{12pt}}rllllc}
\multicolumn{7}{c}{{\bf \large Table 1 }} \\ \\
\multicolumn{7}{c}{{\bf $\mu = 50\; \; $  $\hbar =1\; \;$  $e=1\; \;$
$\gamma_0 = 0.295411 \sqrt \lambda$}} \\  \hline \\

{\bf Type}& {\bf M}&{\bf $\tau$} & {\bf $m$} & {\bf$\lambda$}  & {\bf
$\gamma_0$} & Figure\\ \\
 \hline \\

One Loop & 0.5 & 0.88 & 0.20 &  0.5436 &   0.2178 & {\bf Fig. 6}\\
Two Loop &  &  &  & 0.1454  &   0.1126 &   \\
\\

One Loop & 1 & 0.67 & 0.25 &  0.2226   & 0.1394 & {\bf Fig. 7}\\
Expanded &  & & & 0.5472 &  0.2185 &  \\
Two Loop &  &  &  & 0.0794  &   0.0832 &   \\
\\

One Loop & 5 & 0.42 & 0.55 &  0.0219   & 0.0437 & {\bf Fig. 8}\\
Expanded &  & & & 0.0301 &  0.0512 &  \\
Two Loop &  &  &  & 0.0176    & 0.0391 &   \\
\\

One Loop & 10 & 0.30 & 0.60 &  0.0147   & 0.0358 & {\bf Fig. 9}\\
Expanded &  & & & 0.0146 &  0.0357 &  \\
Two Loop &  &  &  & 0.0114    & 0.0315 &   \\
\\

\\

\end{tabular}

\end{center}

Note that although the shapes of the wells at the true minima are similar for the one loop and two loop potentials, the depths of the wells are quite different. To understand this behaviour we recall that the value
of the
effective potential at the true minimum depends on the
renormalization condition $V|_{v=0}=0$.  The important point is that this condition is imposed at $v=0$ which is in general far from the position of the true minimum. The only physical quantities that are perturbatively reliable are
those
which depend on the properties of the effective potential near the true minimum,
namely $\tau$, $m$ and $\lambda$ (or $\gamma_0$). The first two parameters are fixed
by the renormalization conditions both at one loop and  two loop.  The true test
of the reliability of the perturbative expansion comes from comparing the
values of $\gamma_0$ obtained from (\ref{derconst}) for the one and two loop potentials. Table [1] shows that these values 
compare fairly well, especially for higher values of $M$.

\section{Conclusions}

We have studied spontaneous symmetry breaking for 2+1 dimensional scalar QED
with both  minimal and non-minimal Chern Simons couplings.  The effective
potential for both of these models has been calculated previously
\cite{jeff,ref6}.  Starting from these expressions, we have studied the
renormalization procedure
in an attempt to shed light, perturbatively, on the  nature of the symmetry
breaking
in both models.  All calculations have been carried out using $Mathematica$.

First, we have looked for symmetry breaking at one and two loops in the
minimal model using the physical renormalization point (which means defining the  couplings in
terms of the derivatives of the effective potential at the global minimum).  We have obtained a
renormalized  effective potential that exhibits spontaneous symmetry breaking
(for an appropriate choice of couplings) and is perturbatively reliable.  Our
main results for the minimal model are summarized in Figs. [4] and [5].  Fig. [4] shows that symmetry breaking at one loop  is independent of the choice of the renormalization point.  Fig. [5] shows that two loop corrections near the
minimum are
small. These results disagree with the those of \cite{ref6} in which it was found that the presence
of symmetry breaking at one loop depends on the choice of the renormalization
point. We have shown that this result is a consequence of using a non-physical renormalization point.  

Secondly, we have performed a two loop analysis of the non-minimal model. We use the physical renormalization point and obtain  an effective potential that exhibits spontaneous symmetry breaking (for an appropriate choice of couplings) and is perturbatively reliable. The renormalization
conditions
are considerably more complicated for the non-minimal model, in part because of the more complicated structure of the interaction, and in part because of the fact that all of the derivatives of the effective potential are singular at the origin.  We were not able to obtain a simple dimensional transmutation relation for the non-minimal model.

\acknowledgments

\noindent This work is supported by the Natural Sciences and Engineering
Research
Council of Canada.

\newpage

\begin{figure}
\epsfxsize=8cm
\centerline{\epsfbox{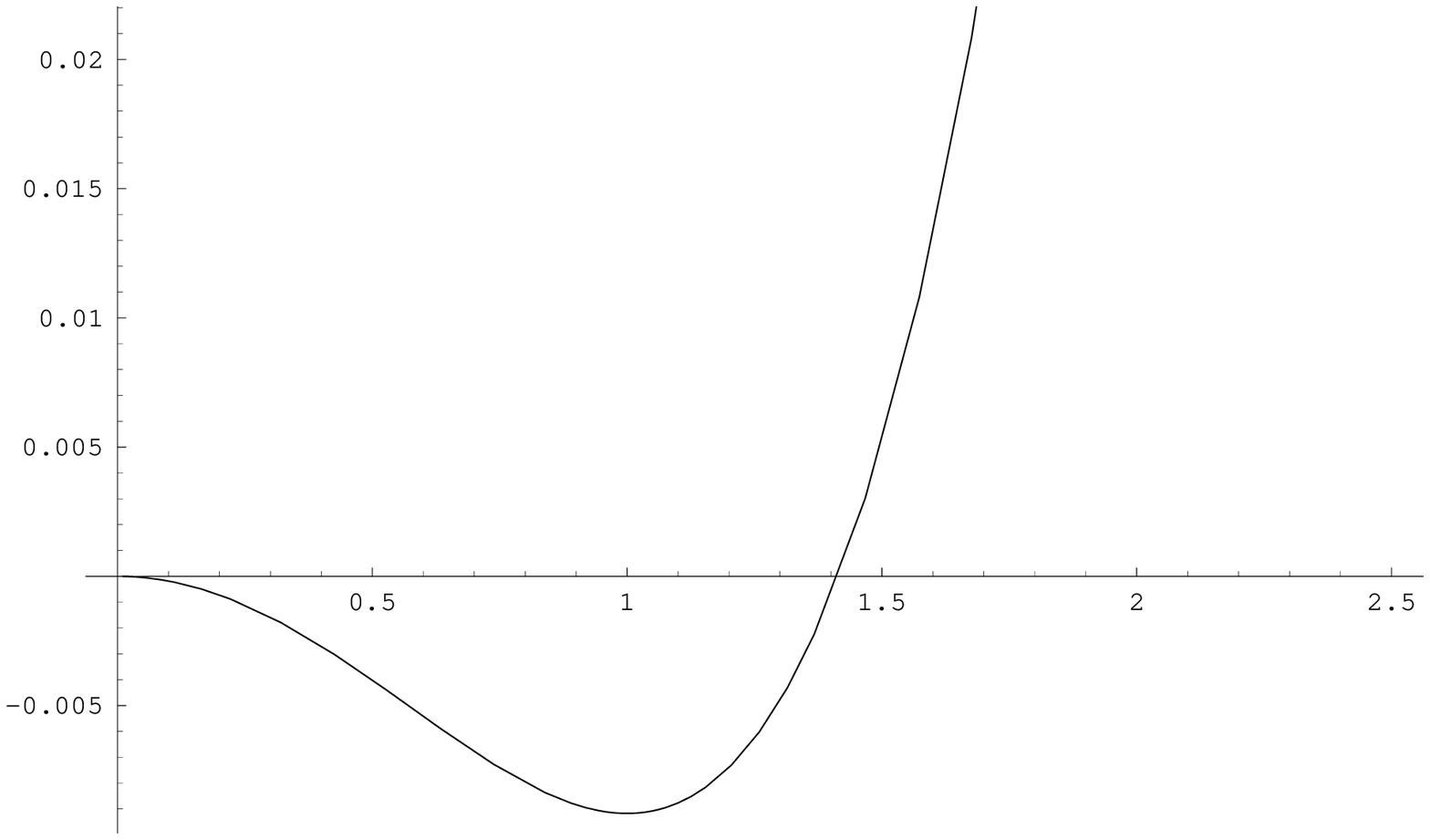}}
\vskip 0.4cm
 \caption{The two loop minimal effective potential using the renormalization
conditions (\ref{rcond2}) and (\ref{derconst}) with $M=1, ~m=.2709$,
$\lambda$=.2200, $\nu$=0.0005 and  $\kappa=20.7494$.}
\label{1}
\end{figure}

\begin{figure}
\epsfxsize=8cm
\centerline{\epsfbox{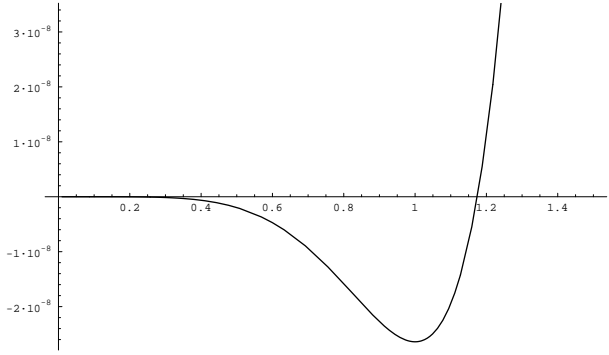}}
\vskip 0.4cm
 \caption{The two loop minimal effective potential using the renormalization
conditions (\ref{rcond1}) and (\ref{derconst}) with $M=1, ~m=\lambda=0$,
$\nu=0.0005$ and  $\kappa=25.1850$.
 }
\label{2}
 \end{figure}
\clearpage
\begin{figure}
\epsfxsize=8cm
\centerline{\epsfbox{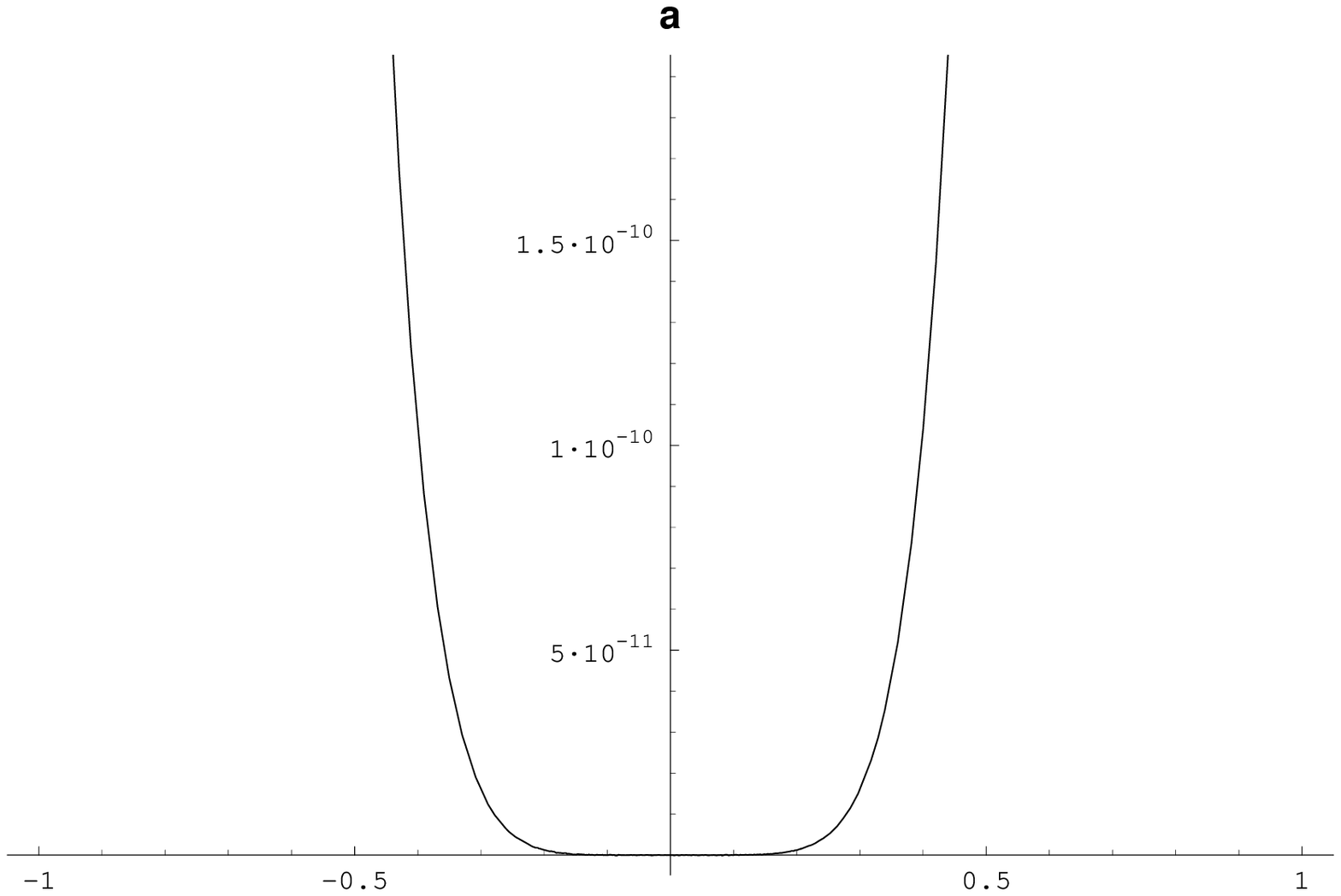}}
\vskip 0.4cm

\epsfxsize=8cm
\centerline{\epsfbox{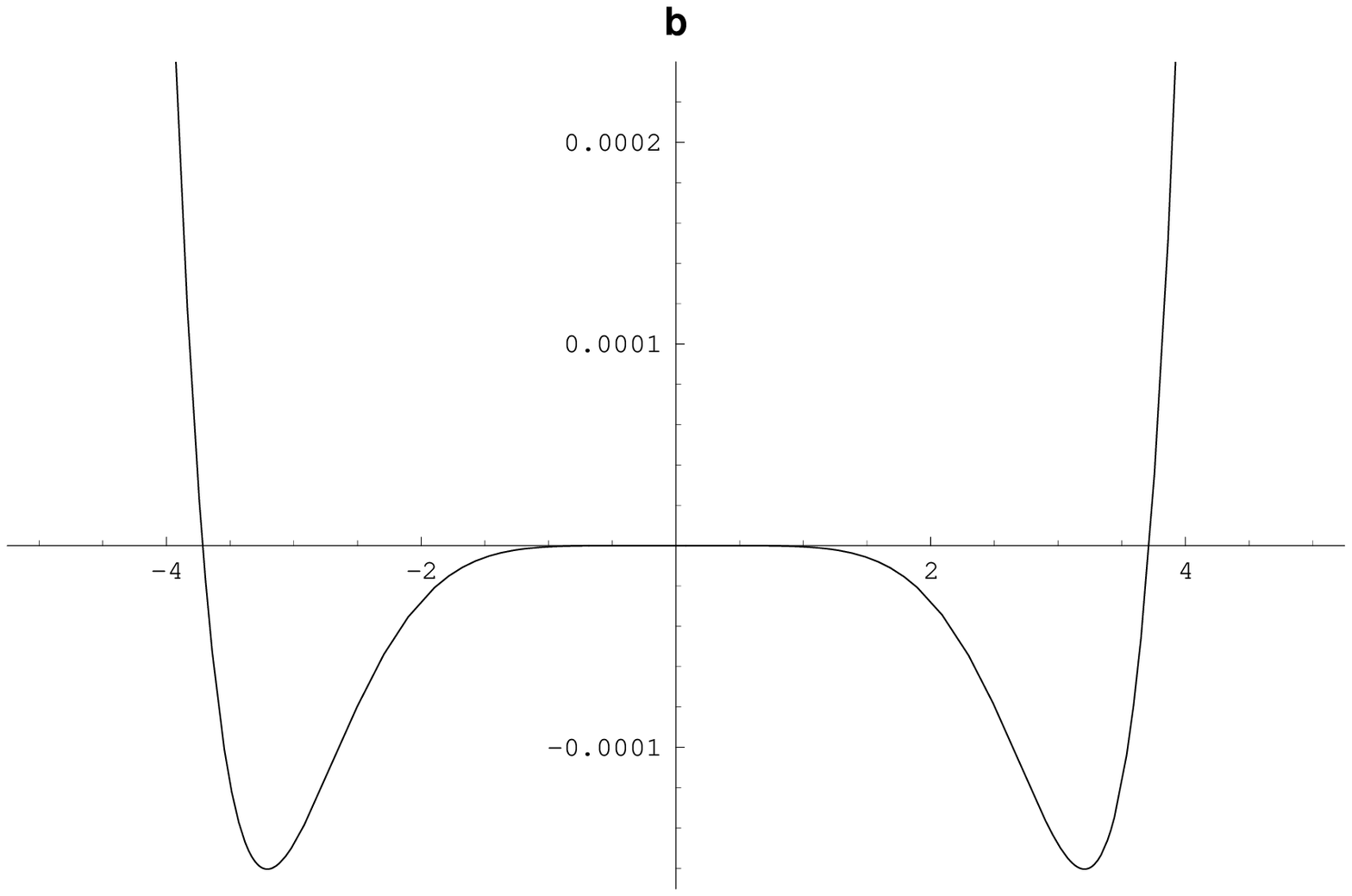}}
\vskip 0.4cm

\epsfxsize=8cm
\centerline{\epsfbox{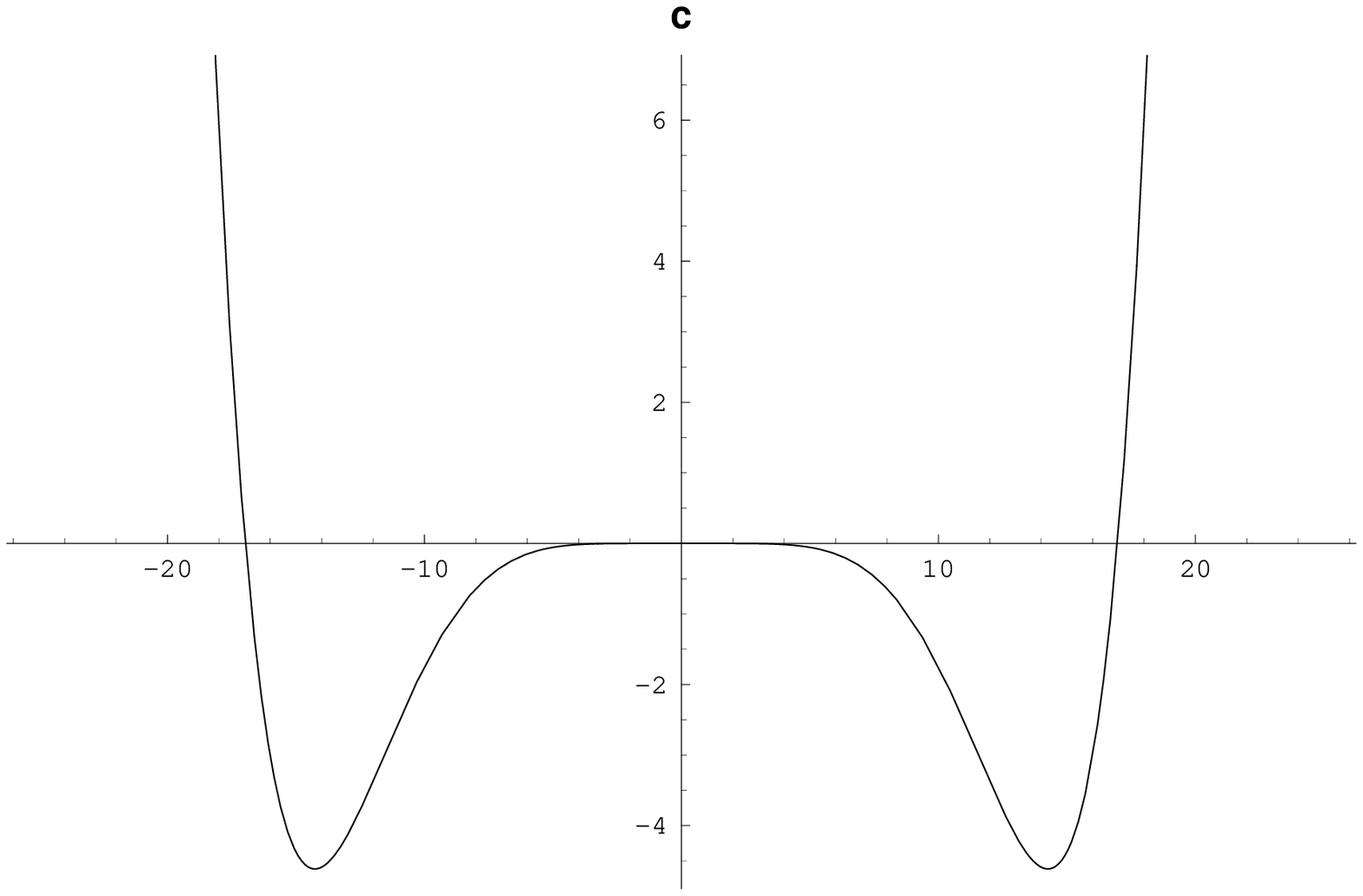}}
\vskip 0.4cm
 \caption{The one loop minimal effective potential using the renormalization
conditions (\ref{rcond1}) with $m = \lambda=0, ~\nu=0.0005, ~\kappa=20$ and (a)
$M = .5$; (b) $M= 1$; (c) $M=5$.  The minima for these graphs are not at
$\sqrt{M}$. There is no symmetry breaking for $M=.5$.}
\label{3c}
 \end{figure}

\begin{figure}
\epsfxsize=8cm
\centerline{\epsfbox{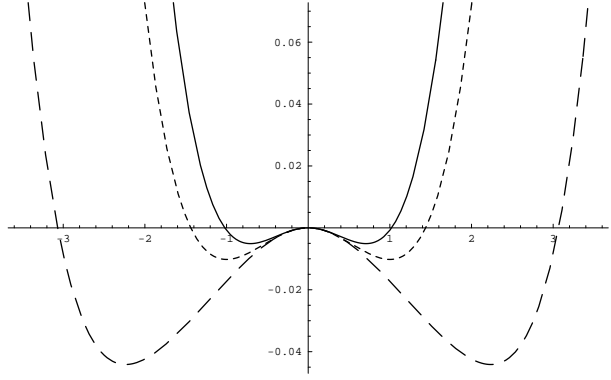}}
\vskip 0.4cm
 \caption{The one loop minimal effective potential using the renormalization
conditions (\ref{rcond2}) and (\ref{derconst}) with $m=.2709, ~\lambda=.2188,~
\nu=0.0005$ and $M=.5$,  $\kappa=1.5588$ (solid line); $M=1$,  $\kappa=1.9748$
(dotted line); $M=5$,  $\kappa=3.4534$ (dashed line). }
\label{4a}
 \end{figure}

\begin{figure}
\epsfxsize=8cm
\centerline{\epsfbox{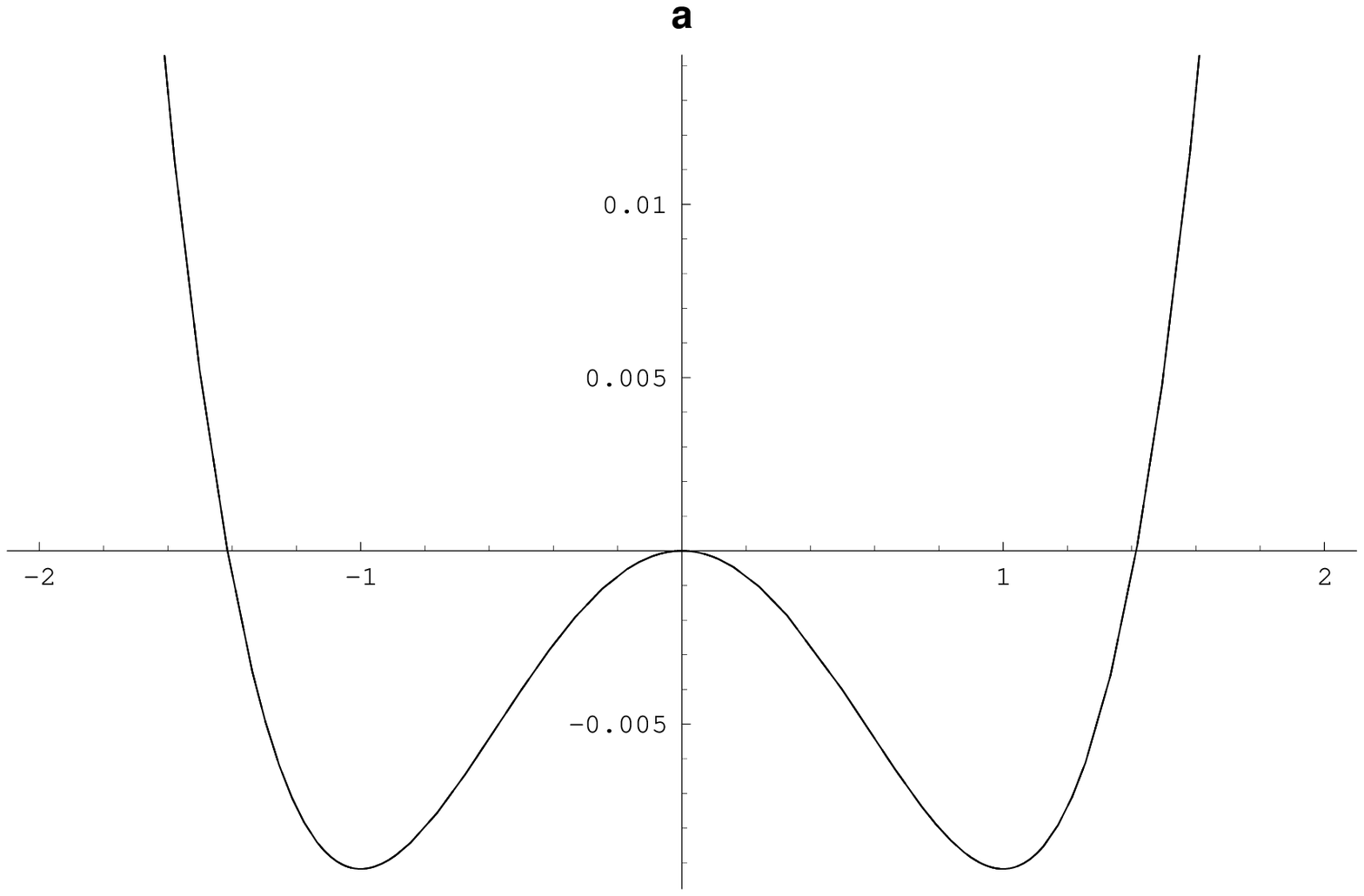}}
\vskip 0.4cm

\epsfxsize=8cm
\centerline{\epsfbox{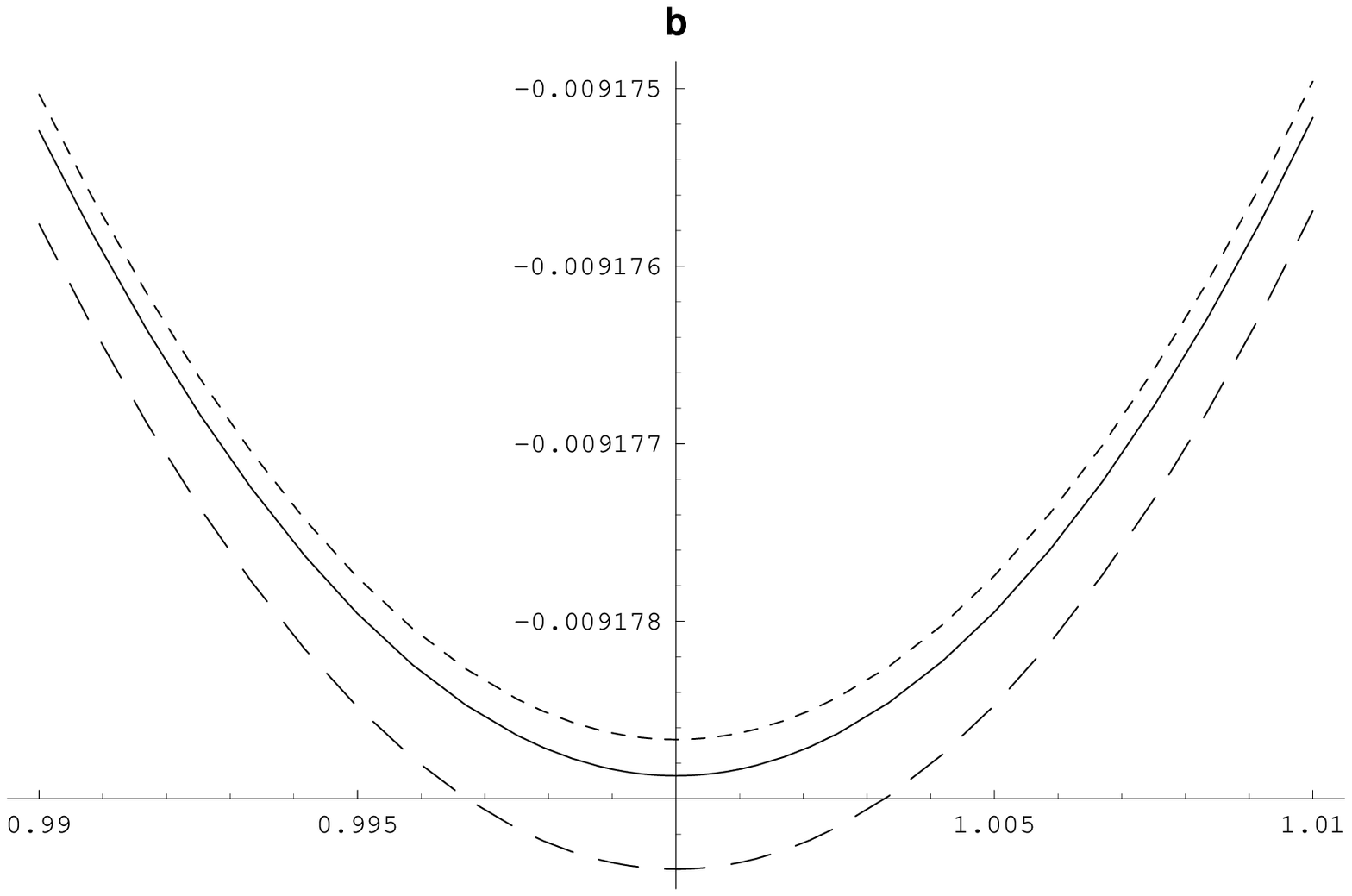}}
\vskip 0.4cm
 \caption{The minimal effective potential using the renormalization conditions
(\ref{rcond2}) and (\ref{derconst}) with $M=1$,  $m=.2709$,  $\lambda=.2188$
and $\nu=0.0005$. The values of $\kappa$ are: one loop $\kappa= 14.2495$
(dotted line); two loop $\kappa$=14.9877 (solid line); expanded $\kappa$ =
10.6287 (dashed line).  }
\label{5b}
 \end{figure}

\begin{figure}
\epsfxsize=8cm
\centerline{\epsfbox{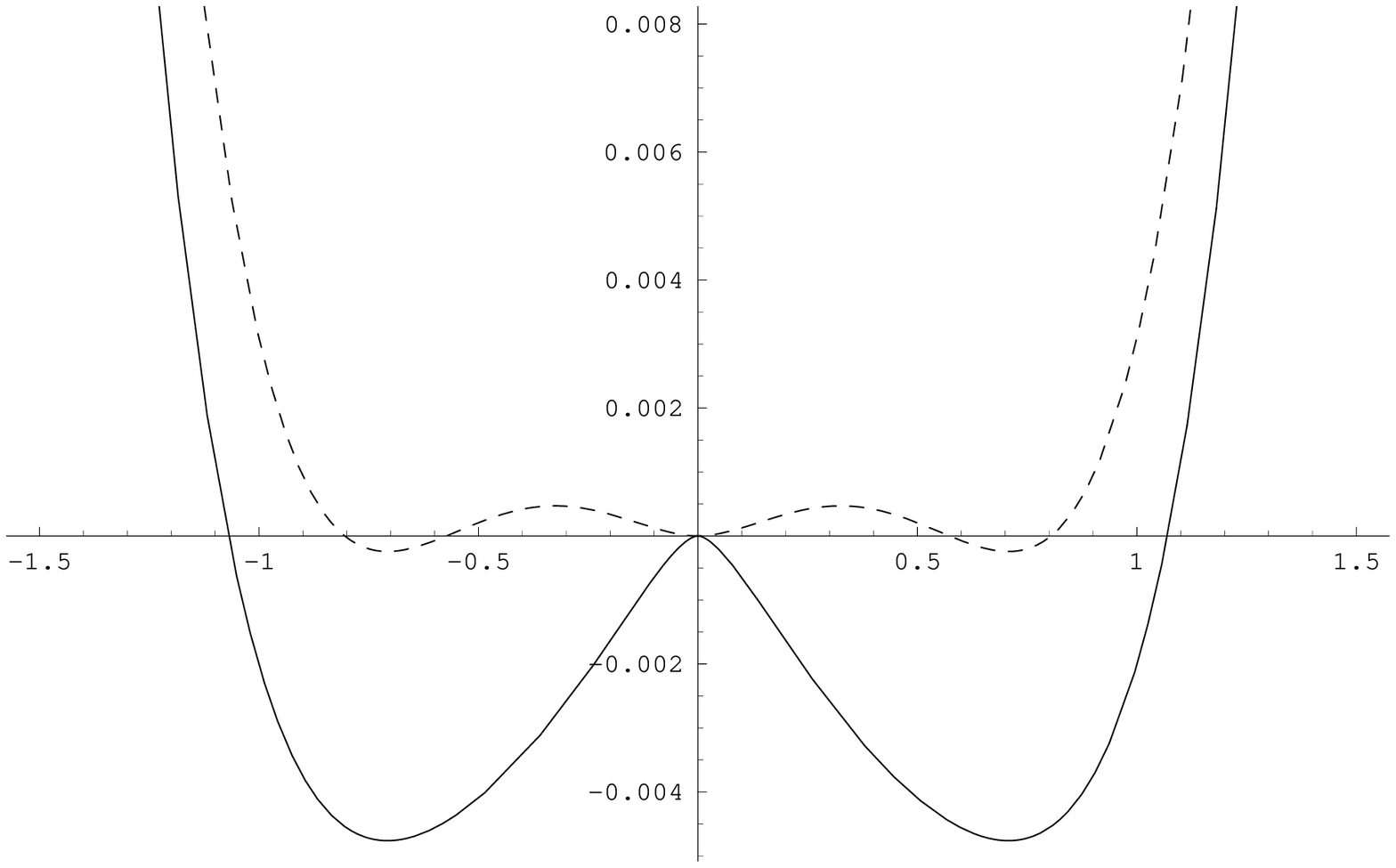}}
\vskip 0.4cm
\caption{Non-minimal effective potential for $M=.5$ and other parameters as
given in Table [1]. The dotted line is the one loop result and the solid
line corresponds to two loops.}
\label{5m}
 \end{figure}

\begin{figure}
\epsfxsize=8cm
\centerline{\epsfbox{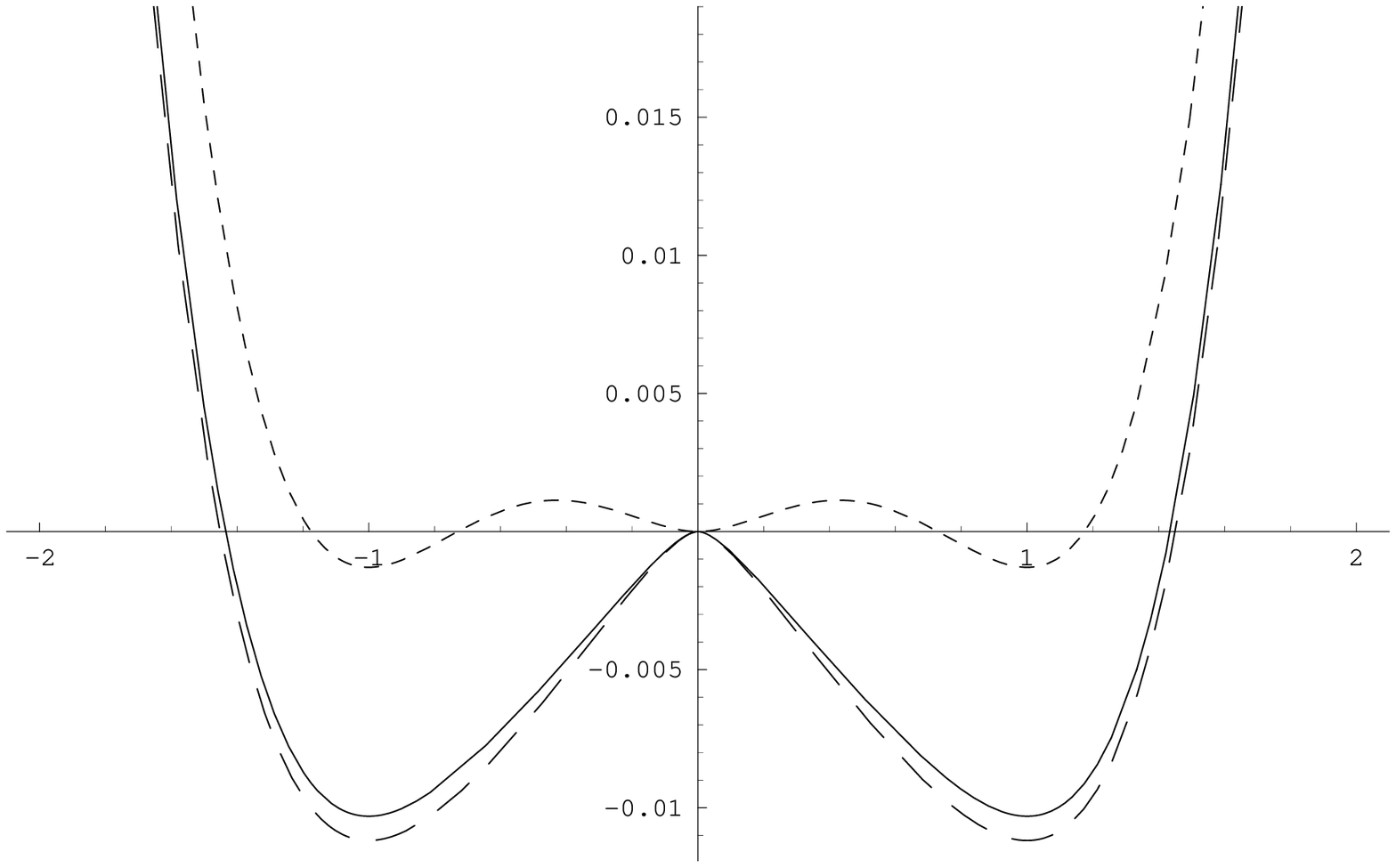}}
\vskip 0.4cm
 \caption{Non-minimal effective potential for $M=1$ and other parameters as
given in Table [1].  The dotted line is the one loop result, the solid
line corresponds to two loops, and the dashed line is the expanded result.}
\label{5i}
 \end{figure}

\begin{figure}
\epsfxsize=8cm
\centerline{\epsfbox{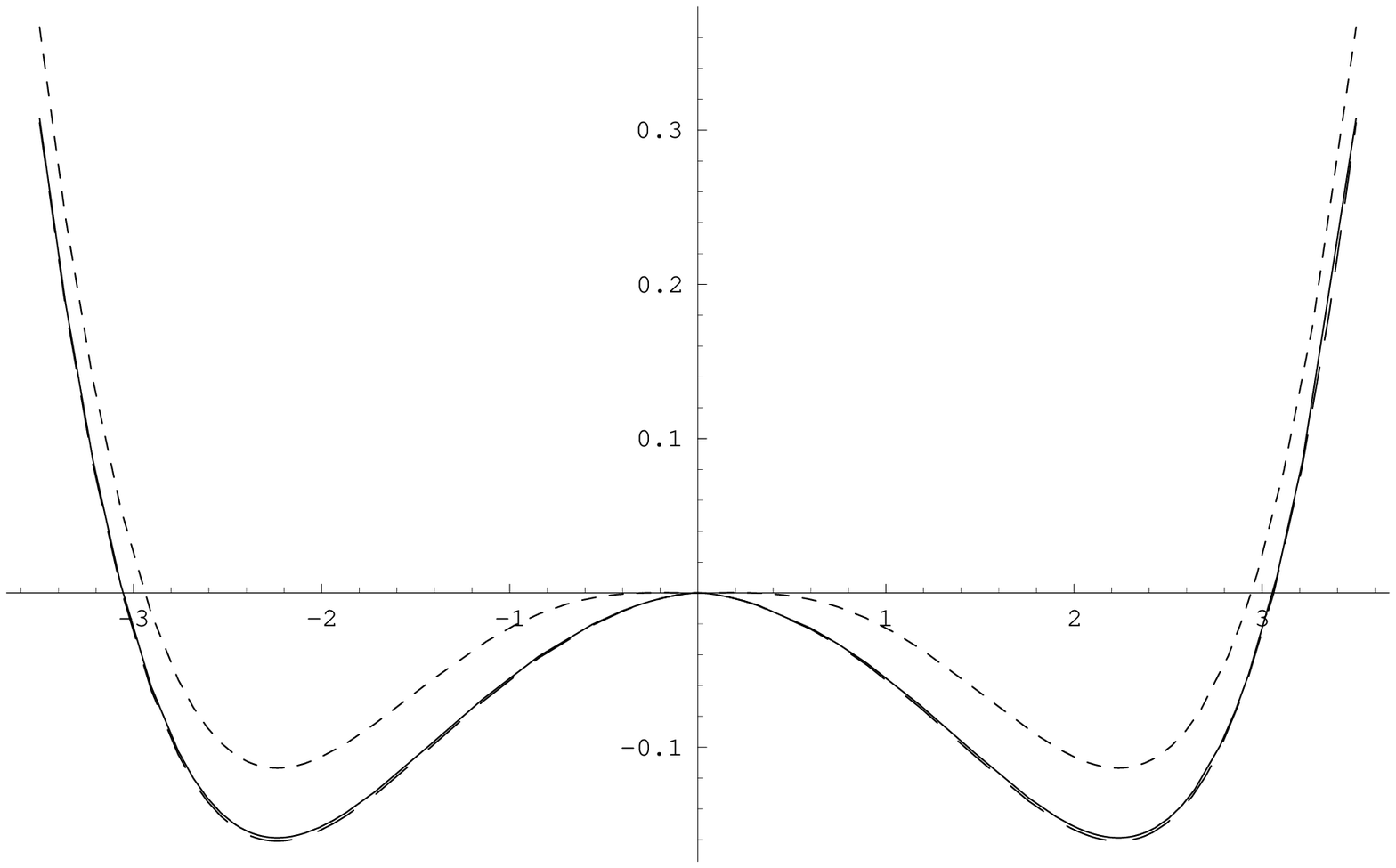}}
\vskip 0.4cm
\caption{Non-minimal effective potential for $M=5$ and other parameters as
given in Table [1].  The dotted line is the one loop result, the solid
line corresponds to two loops, and the dashed line is the expanded result.}
\label{5j}
\end{figure}

\begin{figure}
\epsfxsize=8cm
\centerline{\epsfbox{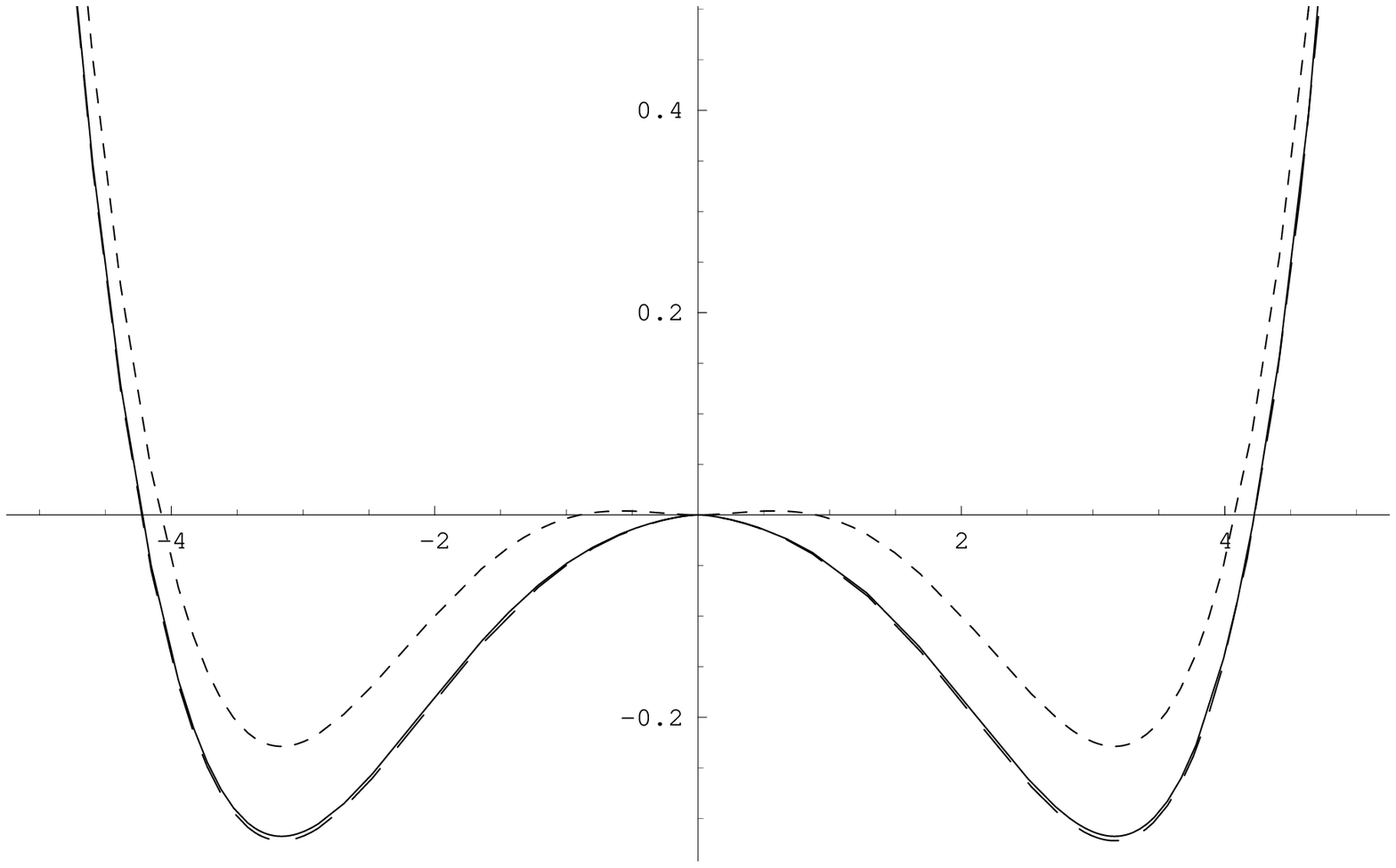}}
\vskip 0.4cm
\caption{Non-minimal effective potential for $M=10$ and other parameters as
given in Table [1].  The dotted line is the one loop result, the solid
line corresponds to two loops, and the dashed line is the expanded result.}
\label{5k}
\end{figure}

\end{document}